\documentclass[aps,preprintnumbers,amsmath,amssymb]{revtex4}
\usepackage{graphicx}
\begin{document}
\title{Nonreciprocal transmission of sound in viscous fluid with asymmetric scatterers}
\author{E.~Walker$^{1,2}$}
\author{A.~Neogi$^1$}
\author{A.~Bozhko$^1$}
\author{J.~Arriaga$^3$}
\author{Hyeonu Hu$^4$}
\author{Jaeyung Ju$^4$
\footnote{Present address: Department of Mechanical Engineering, University of Michigan-Shanghai Jiao Tong University Joint Institute, 800 Dongchuan RD. Minhang District, Shanghai, China }}
\author{A.A.~Krokhin$^1$}
\email{arkady@unt.edu} 
\affiliation{$^1$Department of Physics, University of North Texas, P.O. Box 311427, Denton, TX 76203} 
\affiliation{$^2$Echonovus Inc., 1800 S. Loop 288 STE 396 \#234, Denton, TX 76205} 
\affiliation{$^3$Instituto de F\'{\i}sica, Universidad Aut\'onoma de Puebla Apartado Postal J-48, 72570, Puebla, M\'exico}
\affiliation{$^4$Department of Mechanical \& Energy Engineering, University of North Texas, 3940 North Elm Suite F101, Denton, TX 76207}
\date{\today }

\maketitle

{\bf Two common concepts of nonreciprocity in sound propagation are based on nonlinear effects \cite{Cumm,stat} and on local circulation of fluid \cite{Hab,Yang}. They originate from two known methods of breaking a time reversal symmetry, that is necessary for observation of nonreciprocal effects. Both concepts require additional devices to be installed with their own power sources. Recently it was demonstrated that acoustical losses may serve as a source of {\it T}-symmetry violation, thus leading to nonreciprocity in reflection of sound from gradient-index metasurface \cite{Cumm4}. Here, we explore viscosity of fluid as a natural factor of $T$-symmetry breaking. We report experimental observation of the nonreciprocal transmission of ultrasound through a water-submerged phononic crystal consisting of asymmetric rods. Asymmetry, or broken {\it P}-symmetry, is the second necessary factor for nonreciprocity. Experimental results are in agreement with numerical simulations based on the Navier-Stokes equation. This passive nonreciprocal linear device is cheap, robust and does not require an energy source.}

A source of sound may generate a quite complicated pattern of pressure and velocity in an inhomogeneous medium. The acoustodynamic field can be calculated analytically only for a few simple arrangements of symmetric scatterers. For more complicated geometries, one relies on numerical solutions. In a linear and lossless medium, the accuracy of the solution can be controlled via the reciprocity theorem which states that a signal emitted by a source in a point {\it A} and received in a point {\it B} remains the same if the positions of the emitter and receiver are switched\cite{LL}. This theorem is very general since it originates from the time-reversal symmetry of the dynamical equations for an elastic medium. Therefore, it is valid for anisotropic media, for media with temporal dispersion, and even for media with dissipative losses \cite{Hoop}. At first glance, the latter statement contradicts the general property of irreversibility of any process accompanied by increase of entropy. The following example clarifies this issue. Let an object be moved back and forth along a surface. Because of inevitable friction, this process is irreversible (the entropy increases) but it remains reciprocal if the trajectories of the forward and backward motion are equal. It, however,
becomes nonreciprocal if the trajectories are not exactly the same. 
for the forward and backward motion.

For a wave process, nonreciprocity appears if dissipation changes with the direction of propagation. Dissipation in a viscoelastic medium is usually introduced by adding the imaginary part to the moduli of elasticity. In a medium with complex elastic moduli, a propagating wave decays exponentially but the energy losses accumulated along the "wave trajectory" remain equal for the opposite directions of propagation. This is the physical reason why the reciprocity turns out to be compatible with dissipation. A formal reason of such compatibility is the fact that the wave equation in viscoelastic dissipative media contains only second derivatives over time, i.e. the wave equation is invariant under time reversal. In recent reviews on nonreciprocal propagation of sound \cite{Maz,Fle,Cumm3} as well as in the mathematical proof of the reciprocity theorem \cite{Hoop} the statement regarding the reciprocal propagation in dissipative media is related to the particular class of media with complex elastic moduli.

Complex (or dynamic) elastic moduli are phenomenological parameters which are introduced in the macroscopic approach. A more detailed (microscopic) approach requires calculation of the field of velocities $\bf v(r)$ generated by propagating sound wave. The power dissipated due to viscosity is obtained by integration of the local gradients of velocity \cite{LL}
\begin{equation}
\label{1} \dot{Q}= -\frac{\eta}{2} \int \left(\frac{\partial v_i}{\partial x_k} + \frac{\partial v_k}{\partial x_i} -\frac{2}{3} \delta_{ik} \frac{\partial v_l}{\partial x_l} \right)^2 dV - \xi \int \left( \nabla \cdot {\bf v} \right)^2 dV.
\end{equation}
Here $\eta$ and $\xi$ are the viscosity coefficients. For inhomogeneous fluid they depend on coordinates, $\eta=\eta(\bf r)$ and $\xi=\xi(\bf r)$. The energy losses due to temperature oscillations are omitted in Eq. (\ref{1}) since for water they are negligible.  The vector field of velocities $\bf v(r)$ in a viscous fluid is calculated from the Navier-Stokes equation solved together with the continuity equation \cite{LL}. For weak-amplitude sound waves, these equations can be linearized and the final equation for velocity component $v_i(\bf r)$ takes the following form:
\begin{equation}
\label{2} 
\rho \ddot{v_i}-\frac{\partial}{\partial x_{i}}\left( \lambda \nabla\cdot {\bf v}\right)= \frac{\partial}{\partial x_{k}}\left[ \eta \left(\frac{\partial \dot{v_{i}}}{\partial x_{k}} + \frac{\partial \dot{v_{k}}}{\partial x_{i}} - \frac{2}{3} \delta_{ik} \nabla \cdot {\bf \dot v}\right) \right] + \frac{\partial}{\partial x_{i}} (\xi \nabla \cdot {\bf \dot v}), \,\,\, i = x,y,z.
\end{equation}
Here $\rho=\rho(\bf r)$ is the mass density and $\lambda$ is the bulk elastic modulus of the fluid. For inviscid fluid ($\eta=\xi=0$) this equation is reduced to the wave equation.

Eq. (\ref{2}) is obviously nonreciprocal since all the terms in the right-hand side contain the derivative ${\bf \dot{v}}=\frac{\partial {\bf v}}{\partial t}$, which changes its sign under time reversal. Thus, the reciprocity theorem does not hold for this equation. If the scatterers are asymmetric the vector field $\bf v(r)$ depends on the direction of propagation of sound. Then the same is true for the energy absorbed given by Eq. (\ref{1}). Two sound waves propagating in the opposite directions (for which the inversion symmetry is broken) decay with different rates. This is the source of nonreciprocity in a viscous fluid. The dissipation increases in the regions with strong gradients of velocity. Therefore, scatterers with sharp corners are more siutable for experimental demonstration of nonreciprocity due to gradient induced differential dissipation (GIDD).
\begin{figure}
\includegraphics [width=16cm]{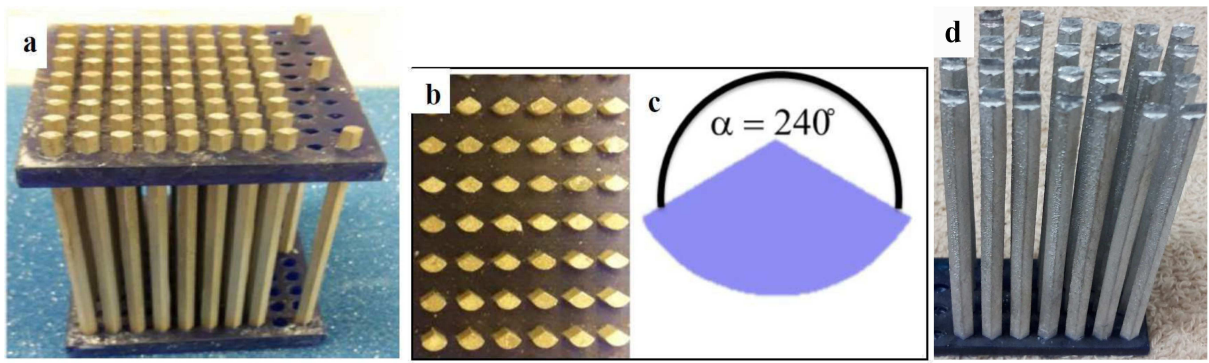}\caption{\textbf{Phononic crystals used for the measurements of acoustic transmission.}  {\bf a}, General
view of the sample with anodized rods. {\bf b}, Top view. The $P$-symmetry is broken along the vertical axis and it holds along the horizontal axis. \textbf{c}, Square unit cell with asymmetric scatterer. The angle $\alpha$ is a measure of broken $P$-symmetry. {\bf d}, Sample with
$4\times 7$ rows of unanodized aluminum rods. }%
\label{fig1}
\end{figure}

For experimental demonstration of nonreciprocal transmission due to GIDD, a phononic crystal of aluminum rods in water environment was used. A sample, shown in Fig. \ref{fig1}, has a square unit cell with  $P$-symmetry broken along the vertical axis. Electrical discharge machining (EDM) was used to fabricate rods out of aluminum that were then anodized to prevent oxidation in water. The period of the lattice is $a=5.5$ mm and the radius of the $120^{\circ}$ circle sector is 2.2 mm. Two V301 1” Panametrics 0.5MHz immersion transducers in a bistatic setup were arranged to measure forward and backward transmission through both the direction with broken $P$-symmetry and without broken $P$-symmetry. Measurements were carried out for two samples with 5 or 7 rows along the direction with broken $P$-symmetry with water at $20^{\circ}$C ambient. Both samples contain 8 rows in the perpendicular direction where the $P$-symmetry holds.  The transmission spectra obtained for these two samples demonstrate similar features, while for the 7-row sample the transmitted signal is expectedly weaker. Here we discuss the results obtained for the 5-row sample.

\begin{figure}
\includegraphics [width=16cm]{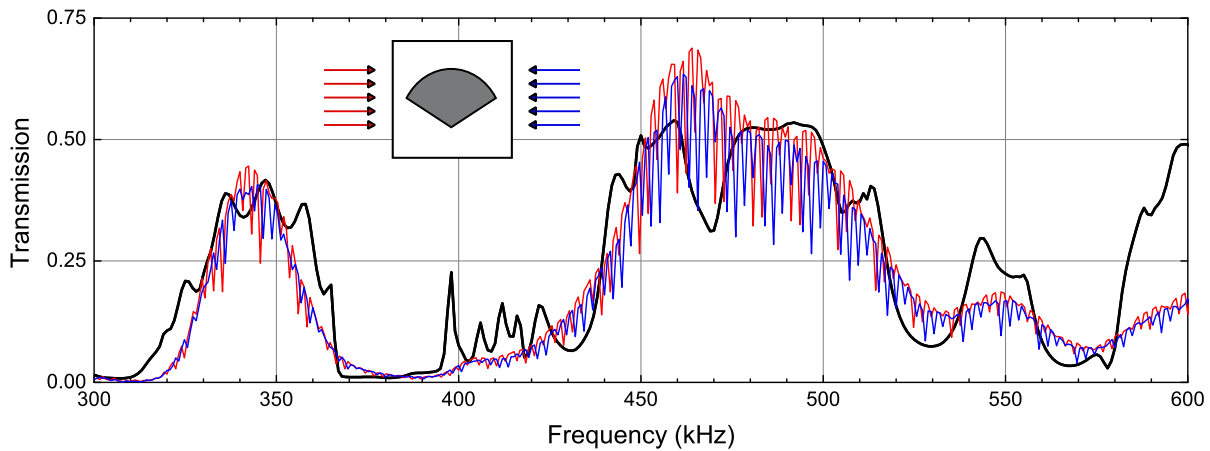}\caption{{\bf{Spectra of reciprocal transmission for the phononic crystal of anodized rods.}} Two measured
spectra of transmission along the direction with $P$-symmetry (red and blue lines) are practically equal, i.e. the transmission is reciprocal. The black line is the transmission spectrum simulated in COMSOL software. Insert shows the orientation of the unit cell with respect to the direction of the incoming wave.} \label{fig2}
\end{figure}

First, the transmission was measured along the symmetric direction (horizontal axis in Fig. \ref{fig1}). The measured spectra for forward and backward transmission are given by two colored lines in Fig. \ref{fig2}. The black line in Fig. \ref{2} shows the transmission spectrum simulated by COMSOL software. Both experimental spectra in Fig. \ref{fig2} show most of the signatures obtained numerically. Because the inverse symmetry, the transmission does not exhibit any regular feature of nonreciprocity. Small fluctuations in the experimental spectra are typical for this type of measurement. While the $T$-symmetry is broken, making the process of propagation irreversible, the reciprocity holds due to the $P$-symmetry. The sound waves experience anisotropic scattering at each rod but they follow the same 'pass' propagating forward and backward. Since the simulated spectra are exactly the same for two opposite directions, only one black line appears in Fig. \ref{fig2}. Thus, the propagation along the direction with $P$-symmetry is reciprocal.

\begin{figure}
\includegraphics [width=16cm]{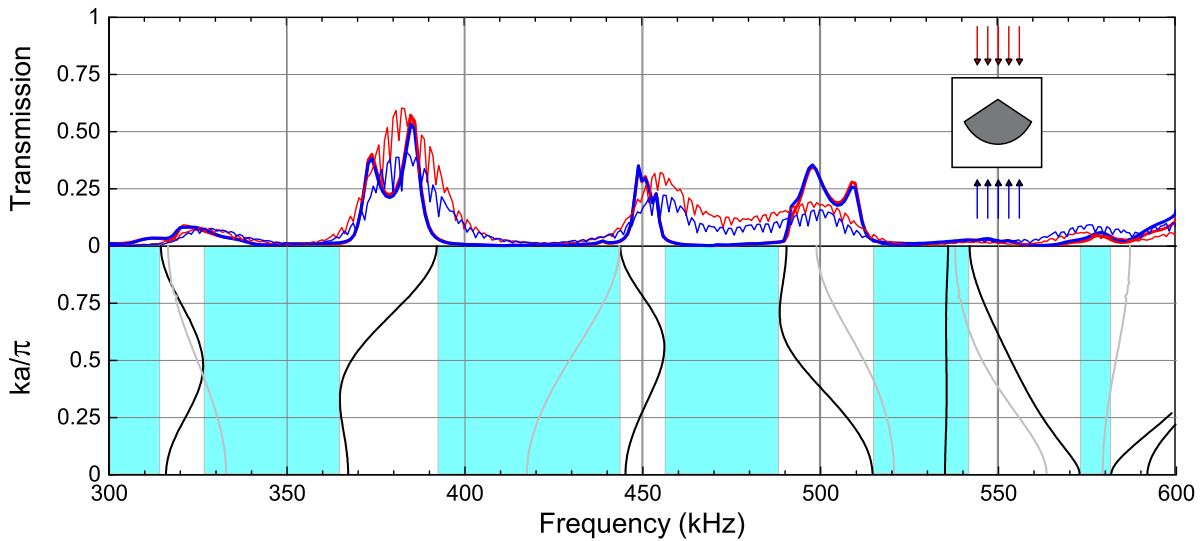}\caption{{\bf Band structure and spectra of nonreciprocal transmission for the phononic crystal of anodized
rods.} {\bf Low panel}, Band structure of phononic crystal with inviscid water background for sound wave propagating along the direction of broken $P$-symmetry. Passing bands corresponding to even (odd) eigenmodes are shown by black (grey) lines. Regions of gaps between the even zones are shaded. {\bf Upper panel}, Wavy lines show experimental spectra for sound waves propagating forward (thin red line) and backward (thin blue line). Numerically calculated transmission spectra are shown by smooth thick lines of the same colors.  Insert shows orientation of the unit cell with respect to the direction of the incoming wave.} \label{fig3}
\end{figure}

\begin{figure}
\includegraphics [width=16cm]{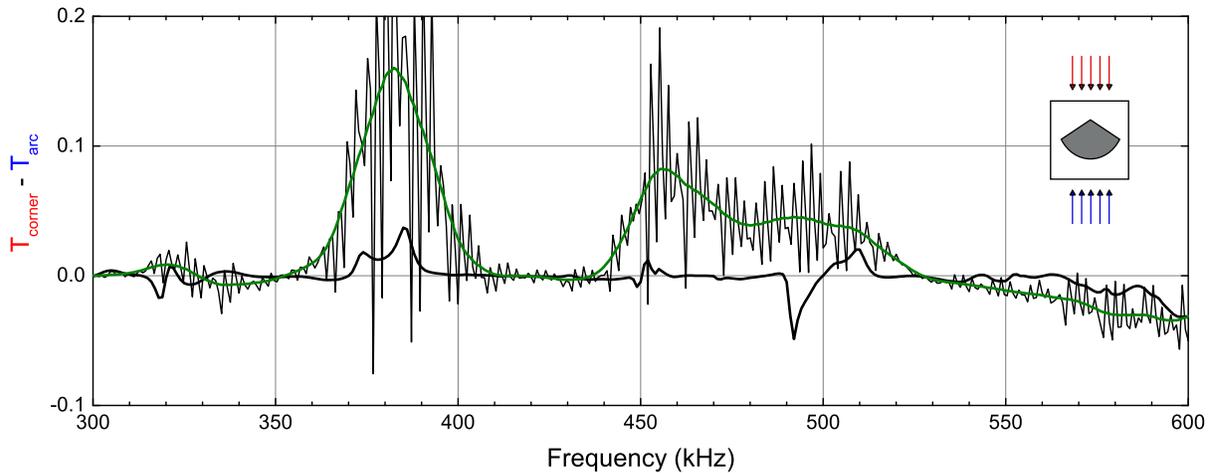}\caption{{\bf The nonreciprocity in the transmission spectra of phononic crystal of anodized rods.} The
difference between the transmissions coefficients plotted in Fig. \ref{fig3}. Experimental (numerical) data are shown by thin (thick) line. Green line is the result of averaging over fast oscillations.} \label{fig4}
\end{figure}

Unlike this, the transmission spectra for propagation along the line of broken $P$-symmetry (vertical axis in Fig. \ref{fig1}) exhibit regular features of nonreciprocity. In Fig. \ref{fig3} experimental (thin lines) and numerical results (thick lines) for the transmission in two opposite directions are plotted together. Theoretical and experimental results are in a reasonable agreement. All the gaps and passing bands of the band structure in Fig. \ref{fig4} are seen in the measured transmission spectra. At normal incidence only the even modes, i.e. the modes that are symmetric over the vertical axis distribution of pressure and velocity, can be excited.  These even modes are shown by black lines in the band structure in Fig. \ref{fig3}. The odd modes turn out to be deaf at normal incidence and they are shown by grey lines. Due to asymmetry of the scatterers, there are relatively large gaps between the even passing bands. They are shaded in Fig. \ref{fig3}. Within the bandgaps the transmission loss reaches upwards of 20 dB. Both experimental and numerical results show relatively high transmission within the gap region for frequencies $326<f<341$ kHz. We attribute this to excitation of the odd eigenmode existing in this frequency range. This becomes possible because the acoustic beam radiated by finite-size vibrating membrane has, of course, some Fourier components with nonzero wave vectors in the horizontal direction. These diffracted components may excite the odd mode. Since the diffraction decreases with frequency, the contribution of the odd higher-frequencies eigenmodes is negligible.

The region where the transmission is higher than theoretically predicted is the bandgap which extends from 456 to 488 kHz. Here the measured transmission exceeds what was numerically simulated. On the contrary, in the next passing band, $488<f<515$ kHz the measured transmission does not reach the simulated value. These two anomalies were observed in all our measurements on the both samples of size  $5\times 8$ and $7\times 8$. We did not find an appropriate explanation for this disagreement. It could be related to the sensitivity of the transducers at the frequencies around 500 kHz. Final transmission within the narrow gap, $573<f<581$ kHz is most likely due to dissipation which smoothes the edges of the gaps and
leads to final density of states within narrow gaps \cite{Hal}. Here, experimental and numerical results agree with each other. 
passing band there are local peaks in the transmission which occur at the frequencies where the group velocity ${\bf V}_g = \partial \omega/
\partial {\bf k}$ vanishes. At these frequencies, which are marked in Figs. \ref{fig3} and \ref{fig4} by bold points, the density of acoustic
states has maxima resulting in the enhanced transmission.

Qualitatively, the nonreciprocity is characterized by the difference $T_{corner} - T_{arc}$ between the acoustic energy transmitted through the phononic crystal when the incoming wave hits the corner (red line in Fig. \ref{fig3}) and the rounded part (blue line) of the rods. This difference is plotted vs frequency in Fig. \ref{fig4}. The experimental curve exhibits fast oscillations which originate from weak irregular fluctuations of the transmitted energy in Fig. \ref{fig3}. The green curve in Fig. \ref{fig4} is obtained by averaging over these oscillations. The nonreciprocity corresponding to the numerically calculated transmission is shown by the black thick line. Nonreciprocity is reduced in the regions of gaps where the transmission is low. The sign of the quantity $T_{corner} - T_{arc}$ changes with frequency because the field of velocities $\bf v(r)$ is formed as a result of interference, which locally may be constructive or destructive. While there is a general agreement between the theory and experiment, some obvious inconsistencies require special attention.

It is clearly seen that the measured nonreciprocity exceeds the numerically simulated one. We attribute this difference to microscopic roughness of the aluminum rods. The rods were anodized to increase their resistance against oxidation in water. It is known that the surface of an anodized sample may have roughness of the size order from a few to dozens of microns. At this scale, the surface of the rods is not flat. Driven by oscillating sound pressure, viscous fluid slows down near the surface of a rod at a typical distance of $\delta = 2 \pi \sqrt{2\eta/(\omega \,\rho)}$. At the frequencies $\omega \sim 10^6$ s$^{-1}$ the thickness of the viscous boundary layer in water (Stokes boundary layer) is estimated to be ideally about a few microns. Since the essential part of acoustic energy dissipates within the boundary layer $\delta$, the micron-size roughness strongly affects the level of dissipation. Roughness not only changes fluid dynamics, it also increases the area where the energy dissipates. Thus, surface roughness increases the dissipation of sound energy that leads to stronger nonreciprocity. Random roughness also can be considered as a stochastic element of the system that breaks the $P$-symmetry at the microscopic level. At the same time, the micron-size roughness does not contribute to scattering because the wavelength of sound in water is about 3-5 mm.

Numerical results predict a deep minimum in the nonreciprocity at the frequency of 492 kHz. This minimum was not observed in the experiment with either  $5\times 8$ or $7\times 8$ samples. Since the experimental spectra in Fig. \ref{fig3} do not strongly reproduce theoretical results, we cannot expect good quantitative agreement for the nonreciprocity data as well.

\begin{figure}
\includegraphics [width=16cm]{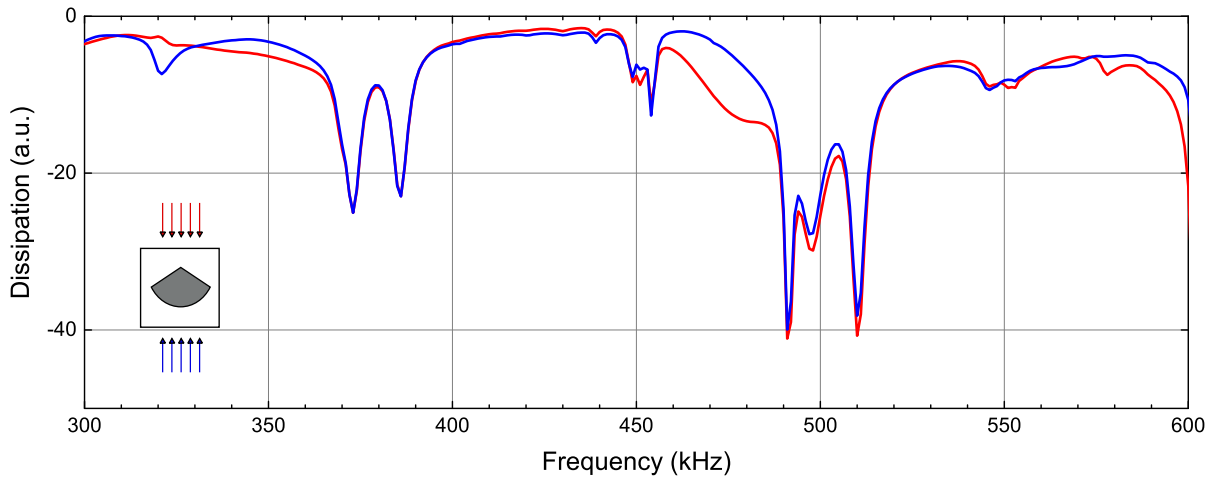}\caption{{\bf Spectra of nonreciprocal dissipation.} Numerically calculated dissipated power for two
opposite directions of sound wave propagating along the direction of broken $P$-symmetry.} \label{fig5}
\end{figure}

The effect of nonreciprocity can be demonstrated not only in the transmitted power, 
ideal (inviscid) fluid,
 but also in the dissipated power given by Eq. (\ref{1}). The distribution of velocities $\bf v(r)$ was calculated for a set of frequencies from
 300 to 600 kHz. Near the rods, the size of the evaluation cell for simulation must be less than one micron in order to resolve fast drop of
 velocity within the narrow boundary layer. Away from the rods, the velocity field changes at much greater distances, therefore the size of the
 mesh cell can be considerably larger. The gradients of all components of velocity were calculated over the region occupied by $5 \times 8$ sample
 and the integral (\ref{1}) was calculated for each frequency. The result of these calculations is presented in Fig. \ref{fig5}.  The difference
 between the graphs corresponding to the opposite direction of propagation is due to the nonreciprocity induced by broken $P$-symmetry in viscous
 water. Due to the high quality of the anodized rods, the dissipated energy is a small fraction of the total energy of the wave. The attenuation
 of the signal is mostly determined by the amount of reflected energy but not by dissipation.  Therefore, higher level of dissipation in Fig.
 \ref{fig5} does not necessary corresponds to lower transmission in Fig. \ref{fig3}. In a homogeneous fluid, the viscosity losses \cite{LL} grow
 quadratically with frequency, $\dot Q \sim (4\eta/3 + \xi) \omega^2$. In a periodic structure, the amount of dissipated energy is affected by
 interference of waves. The frequency dependence $\dot Q(\omega)$ in Fig. \ref{fig5} is a nonmonotonic function, which is strongly suppressed
 within the band gaps.

Transmission is always related to the size of the sample. If the medium is absorbing, it is meaningless to consider samples with length much exceeding the decay length. If the violation of $T$-symmetry is related to dissipation, the decay length of sound limits the size of the sample. In this sense, nonreciprocity due to GIDD is a size-effect and the results are size-dependent.

Transmission spectra shown in Fig. \ref{fig2} exhibit no reciprocity since along horizontal direction of propagation the lattice has inverse symmetry. This symmetry is strongly broken along the vertical direction of propagating wave, therefore the spectra in Fig. \ref{fig3} exhibit maximal nonreciprocal transmission. For any oblique direction the {\it P}-symmetry is broken with some intermediate level. This will lead to the nonreciprocal transmission but with contrast less than that shown in Fig. \ref{fig3}. Similar nonreciprocal effects related to orientation of the sample with respect to incident wave was reported in Ref. [\onlinecite{Cumm4}].

While the energy dissipated in the sample is small, it far exceeds the energy dissipated within equal volume of free water. The viscous decay length of 300 kHz sound in free water is about 100 m. Sound waves propagating through a phononic crystal decays much faster. This occurs due to multiple reflections from solid surfaces of the rods. Each reflection is accompanied by high absorption\cite{LL} with the rate $\sim \sqrt{\eta \omega}$ . Dissipation of acoustic energy in a phononic crystal can be calculated using perturbation theory over the terms proportional to the viscosity coefficients in Eq. (\ref{1}). After quite long calculations the linear correction $\Delta \omega_n(\omega,\bf k)$ to the $n$th eigenfrequency $\omega_n(\bf k)$ of a lossless phononic crystal can be expressed through multiple sums over reciprocal lattice vectors \cite{supp}.  This correction has real and imaginary parts which describe the frequency shift and decay of the $n$th eigenstate. Evaluating the imaginary part near the frequency $f_0=373$ kHz, where $|\dot Q|$ has a local maximum, we obtained that the decay length ${1/\text{Im}k}={|V_g(f_0)|}/{\text{Im} \Delta \omega_2(f_0)}$ does not exceed 10 m. Here ${\bf V}_g=\partial \omega /\partial {\bf k}$ is the group velocity.  This decay length is one order of magnitude less than that in free water. Really, the decay length is probably even less due to surface roughness, which cannot be taken into account in these calculations. Decrease of the decay length due to presence of solid scatterers was predicted in Ref. [\onlinecite{Edgar}], where the effective viscosity has been introduced in the long-wavelength limit. Calculated correction $\Delta \omega_n(\omega,\bf k)$ opens a way to introduce the effective viscosity for 2D phononic crystal at any frequency.

\begin{figure}
\includegraphics [width=16cm]{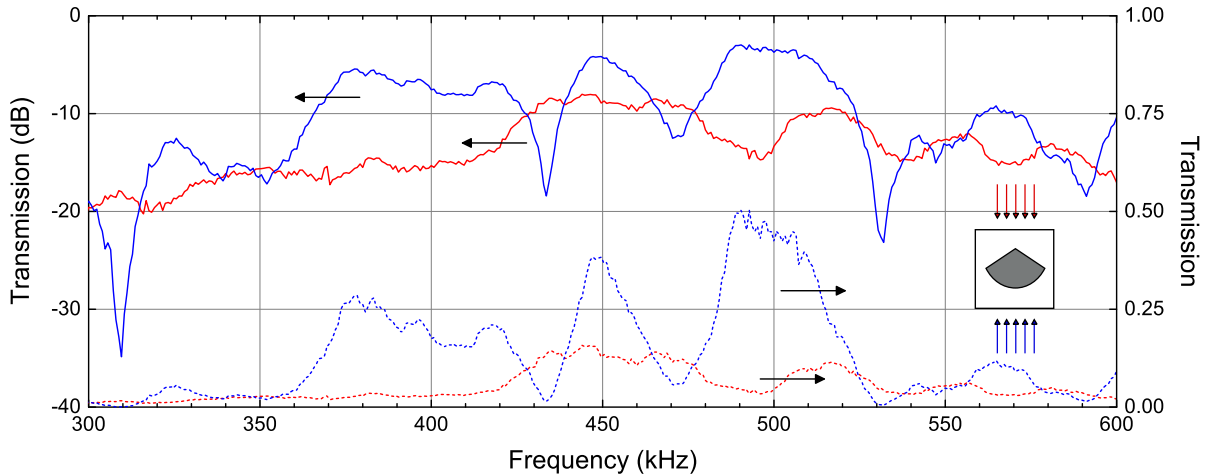}\caption{{\bf The transmission spectra of the phononic crystal of unanodized rods shown in Fig \ref{fig1}
(d).} Solid lines show the transmission plot in logarithmic scale with left vertical axis. Dotted lines show the linear transmission with right vertical axis. The nonreciprocity level is about 10-15 dB within the transmission bands. Insert shows orientation of the unit cell with respect to the direction of the incoming wave.} \label{fig6}
\end{figure}

The nonreciprocity in the spectra shown in Fig. \ref{fig3} is quite weak, achieving a maximum of about 5 dB. It cannot be strong because it originates from the difference between two quantities (acoustic absorption) and each one is weak by itself. Thus, stronger nonreciprocity require higher level of viscous dissipation. The latter can be increased not only by increasing the viscosity of the background fluid but also by using rods with rougher surfaces. Roughness increases the effective area of the rods where acoustic energy is dissipated. To demonstrate stronger nonreciprocity, we used a phononic crystal with the same parameters as shown in Fig. \ref{fig1} (a)-(c), where anodized aluminum rods are replaced by unanodized rods, see Fig. \ref{fig1} (d). These unanodized rods were formed using investment casting in a mold and their surfaces are of much lower quality than that of anodized rods. Transmission spectra for propagation along the direction of broken $P$-symmetry is shown in Fig. \ref{fig6} in linear and logarithmic scale. For this shorter ($4\times 8$) sample the details of the band structure are not well-manifested because of much stronger dissipation. Here the nonreciprocity reaches 10-15 dB, i.e. it is much stronger than was observed for the anodized sample. The wave that propagates towards the sharp corner of the rods is strongly suppressed as compared to the reversed wave. Such level of nonreciprocity allows rectification of acoustic signals, while it still remains lower than that reported in Refs. \cite{Cumm,Hab} where acoustic nonreciprocity was achieved by either nonlinearity or by air-flow bias. An important advantage of the proposed device is the broadness of the band of nonreciprocal transmission. It turns out to be orders of magnitude wider than the band of nonreciprocal transmission of the earlier reported devices.

The nonreciprocal transmission shown in Figs. \ref{fig3}, \ref{fig4}, \ref{fig5} and \ref{fig6} is due to simultaneously broken $T$- and $P$-symmetries. If only $P$-symmetry is broken, (usually due to asymmetry of the scatterers) the effect of different transmission in the opposite directions reported in Refs. \cite{Kris,Li,He} is of geometrical nature. Being pure reciprocal \cite{Maz}, it, nevertheless, leads to unidirectional transmission and thus can be successfully used for rectification of sound.

True nonreciprocal transmission requires violation of time-reversal symmetry. In the present study, it was done by including the viscosity-dependent terms in the dynamical equations. Even this may not be sufficient if the forward and backward waves produce the same distribution of pressure in space. That is why the inversion symmetry was broken by introducing asymmetric scatterers. In the case of layered 1D systems, the asymmetry is irrelevant and observation of truly nonreciprocal transmission requires special efforts. Recently an original 1D system exhibiting nonreciprocal transmission through three layers -- compensated lossy and gain layers separated by a nonlinear layer -- was reported in Ref. [\onlinecite{Gu}].


Nonreciprocity induced by broken $PT$-symmetry combined together with already existing acoustic hyperbolic metamaterials \cite{Cumm2} leads to the concept of a new class of metamaterials. For these metamaterials nonreciprocity goes far beyond nonequal transmission, giving rise to different dispersion for the waves propagating in the opposite directions. Let a finite-size sample of hyperbolic metamaterial have a topological transition at some frequency $\omega_c$. For $\omega < \omega_c$ the dispersion is hyperbolic and it is elliptic for $\omega > \omega_c$. If viscosity is neglected, the transition frequency $\omega_c$ is independent of the direction of propagation. However, in a viscous medium the eigenfrequencies acquire a complex direction-dependent correction $\Delta \omega_n$, which leads to the frequency shift and damping of the eigenmodes. The imaginary part of $\Delta \omega_n$ (viscous damping proportional to $\dot Q$) is different from zero for any shape of the rods. The real part (frequency shift) vanishes for rods with inverse symmetry. It is different from zero only if the {\it PT}-symmetry is broken\cite{supp}.  Unlike the imaginary part of the correction which is always negative, the shift can be either positive (blue) or negative (red). If for $\omega_c$ the shift is positive for forward direction and negative for the backward direction the wave with frequency $\omega_c$ will exhibit respectively elliptic or hyperbolic dispersion, depending on the direction of propagation. Realization of this scenario requires very detailed design of the structure since many different (and contradictory) conditions must be satisfied. For example, the dissipation must be strong enough in order to provide sufficient level of nonreciprocity, but the shift must be greater than the decay in order to overcome the broadening of the topological transition induced by dissipation. While there may be essential technical difficulties in the way of design and fabrication of nonreciprocal hyperbolic metamaterials, the proposed physical principle opens new possibilities to manipulate sound waves.

In conclusion, a new mechanism of nonreciprocal acoustic transmission through a medium with broken $PT$-symmetry is presented. Since the violation of time-reversal symmetry is due to finite viscosity, propagation of sound is described by Navier-Stokes equation. Unlike widely-used approach where dissipation is introduced through complex elastic moduli, viscous fluid dynamics leads to truly nonreciprocal propagation of sound if inversion symmetry is broken. The proposed mechanism can be observed using a passive linear device -- phononic crystals with asymmetric scatterers. Using this passive device, which does not require an external source of energy, nonreciprocity is observed within very wide ranges of frequencies. It is demonstrated that the level of nonreciprocity increases for scatterers with rough surfaces that means that the effective viscosity can be tuned by changing the quality of the surface of the scatterers.

{\bf Materials and Methods} \\

The anodized GIDD rods used in this work were fabricated using conventional electrical discharge machining (EDM). Initially, H13 grade raw aluminum was pre-machined into cylindrical rods of the desired height and radius of 2.2 mm. A CAD design of the GIDD rods consisting of a $120^{\circ}$ section of the 2.2 mm rods was programmed into a Mitsubishi FA10S EDM which then cut the rods to the desired shape. The machined jobs were inspected for the specified dimensions and any visible EDM wire marks were removed by an emery sheet.

{\it Anodizing of GIDD rods}: Initially wire-cut jobs were pre-cleaned and chocked in alkaline solvents for 10 minutes. The rods were then cleaned in de-ionized water and neutralized by chocking in Artec chemical (Nitric acid) for two to five minutes. The jobs were again cleaned using de-ionized water for two minutes. And now the jobs are moved to traditional anodizing process where required color of choice can be obtained. Finally, anodized jobs are cleaned with deionized water and left aside for drying.

\end{document}